\documentclass[aps,prd,groupedaddress,preprint]{revtex4}

\usepackage{graphicx}

\begin{document}

\title{The 2+1 Dimensional $Q\bar{Q}$ Potential in Fourth Order
Color-Dielectric Transverse Lattice QCD}

\author{Bob F. Klindworth}
\author{Matthew Bierman}
\author{Adam Bush}
\affiliation{University of Wisconsin-La Crosse\\La Crosse, WI 54601}

\date{\today}

\begin{abstract}
The 2+1 dimensional (2 space dimensions and 1 time dimension)
Transverse Lattice is constructed using the Color Dielectric formalism.
The dynamics of the Color Dielectric link fields are not known from first
principles, so an ansatz for the effective action is made, which
is truncated at fourth order in the fields. Since the Transverse
Lattice is a Light-Front field theory, it lacks manifest rotational
invariance. This lack of manifest rotational invariance is exploited
to determine the couplings in the effective action by calculating the
$Q\bar{Q}$ Potential, an observable that is very sensitive to rotational
symmetry. This calculation lays the groundwork for a fully 3+1
dimensional treatment of the Transverse Lattice.
\end{abstract}

\pacs{}

\maketitle

\section{Introduction}
Theory and phenomenology both have an important place in helping us
understand the nature of the strong interaction. The very best phenomenological
models strip away the superfluous aspects of the theory and focus solely on
the mechanisms which underly the physics at hand. Rigorous theory, however,
provides the crucial framework upon which our basic physical understanding
rests. In order to perform specific calculations, however, approximations
must be made to the rigorous theory of the strong interaction.
Choosing what approximations to make and determining the affect of
these approximations are some of the challenges in
performing such calculations. \\

The Color Dielectric Formulation of the Tranverse Lattice provides a
means of bridging the gap between rigorous theory and phenomenology. The
Transverse Lattice is a Light Front field theory in which the Light
Front Coordinates, $x^\pm=\frac{1}{\sqrt{2}} \left ( x^0 \pm x^3 \right )$
\cite{dirac},
are continuous variables
and the remaining two space-time coordinates, $x^{1,2}$, are discrete
quantities. The longitudinal dynamics ( the $x^\pm$ directions) are
provided by an instantaneous interaction established by t' Hooft 
in the limit of a large number of quark colors, $N_c$, in 1+1 
dimensional theories\cite{thooft}.
The transverse dynamics are mediated by link fields
which share some similarity to the link fields in Euclidean Lattice
simulations. An important difference is that these link fields are not
the localized gluon fields developed in the usual Euclidean
Lattice framework. Instead, they represent an average over a number
of these localized gluon fields. It is for this reason that such a
formulation of the Transverse Lattice is labeled a Color Dielectric
formulation. \\

The Color Dielectric formulation is analogous to the dielectric
approximation in electromagnetic materials where the electromagnetic
field is smoothed by taking its spatial average. The localized
gluonic fields, like the microscopic electromagnetic field in a dielectric, 
can vary significantly over short distance scales.
Color Dielectric theory averages over many of
these fields, and these new averaged link fields are the effective degrees
of freedom of the theory.
If the average is a spatial average
the link fields can account for (relatively) long distance physics using
comparatively few degrees of freedom.\\

In this paper, the Transverse Lattice is used to calculate
the static $Q\bar{Q}$ Potential in 2+1 dimensions. The dynamics of the
Color Dielectric link fields are not known from first principles, so an ansatz
for the interactions of the link fields is made with the couplings left
as free parameters. These parameters are tuned to reproduce the rotational 
symmetry of the $Q\bar{Q}$ Potential. In this way, the dynamics of
the Color Dielectric link fields can be determined.

\section{Formulating the Transverse Lattice Hamiltonian}
The Transverse Lattice was first formulated by Bardeen and Pearson 
\cite{bardeen1,bardeen2}.
In those papers,
the hamiltonian is formulated in the limit of a large number
of quark colors. This has the effect of limiting the number of terms which
contribute to the effective hamiltonian. The large-$N_c$ limit also provides
a means of systematically excluding quark pair production since quark
pair production mechanisms arise from non-planar diagrams. 
In \cite{us} the effective action 
was truncated at second order in the link fields. Although the results were
acceptable for a first study, there are two important reasons to examine
terms which are fourth order in the link fields. The first reason is obvious:
including fourth order terms is less retrictive. In an ideal world, we would
allow all orders in the link fields to contribute. Truncating the expression
for the effective action is necessary to do practical computations, but
the later we truncate, the closer we will be
to an ``exact'' hamiltonian. The second reason is that fourth order terms
are necessary to model 3+1 dimensions. The canonical plaquette interaction
is the simplest interaction which distinguishes two link fields strung
end to end in a straight line versus two link fields end to end going around
a corner on the lattice. The plaquette interaction term is fourth order in the
link fields. For this reason, and for consistency, we choose to include the
fourth order terms which contribute in our approximation. These terms
are represented by,

\begin{eqnarray}
H^{(4)} & = &\lambda_1 Tr\left [M_{i,j,x}M_{i+1,j,x}M_{i+1,j,x}^{\dagger}
M_{i,j,x}^{\dagger} \right ] \nonumber\\
& + &\lambda_2 Tr\left [ M_{i,j,x}M_{i+1,j,y}M_{i+1,j, y}^{\dagger}
M_{i,j,x}^{\dagger}\right ] \nonumber\\
& + &\lambda_p Tr\left [ M_{i,j,x}M_{i+1,j,y}M_{i+1,j+1,x}^{\dagger}
M_{i,j,y}^{\dagger}\right ]
\end{eqnarray}

The first term represents the interaction between two link fields laid end to
end in a straight line. The second term accounts for the interaction
between two link fields which form a corner. The final term is the
canonical plaquette interaction with the link fields being multiplied in order
around a simple plaquette in the lattice. 
In addition to these fourth order interactions, a ``contact'' interaction was
introduced which acts only on the links at the ends of the ``string''
connecting the quarks. The coupling associated with this interaction is
labeled $\lambda_c$.
The couplings $\lambda_1$,
$\lambda_2$, $\lambda_p$, and $\lambda_c$
are free parameters in the theory which must
be determined. They are free parameters since the detailed dynamics of
the effective degrees of freedom represented by the link fields are not
known from first principles. In \cite{dalley}, Lorentz symmetry was used
as the benchmark for tuning the couplings.
As we shall see, another excellent observable with which to tune these
parameters is the static $Q\bar{Q}$ Potential.

\begin{figure}
\includegraphics[scale=0.50]{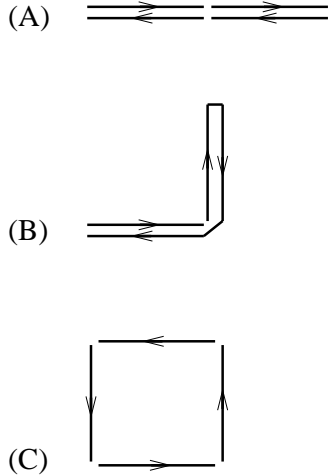}
\caption{Schematic view of the terms in the effective action. (A)
represents the link fields in a straight line interacting,
(B) represents the link fields
around a corner interacting, and (C) is the plaquette interaction.
\label{cartoon}}
\end{figure}

\section{Calculating the $Q\bar{Q}$ Potential}
The $Q\bar{Q}$ Potential is calculated using the same method as in our
previous work \cite{us}. The Discrete Light Cone
Quantization (DLCQ) formalism is exploited for the transverse 
directions \cite{dlcq, dlcq2} . The hamiltonian is constructed for a Fock space
which has two essential approximations. The first is that only Fock states with
total momentum less than some cut-off are considered. This approximation
is necessary to give the problem a finite size, making it amenable
to computational solution. The second approximation is the {\it minimal
string approximation} in which the color sources are connected by a
link field string of the shortest possible length in lattice units. This
is an {\it ad hoc} approximation, however, in hindsight it appears to
not affect the results of the calculation to any great degree. For
a system with two link fields between the quarks, the eigenvalue equation
is found to be,\\

\begin{eqnarray*}
P^-\psi(p_1,p_2) = \left ( \frac{p_1 + p_2}{2v^{+\ 2}} +
\frac{\mu^2}{2p_1} + \frac{\mu^2}{2p_2} \right )\psi(p_1,p_2)\\
+ G^2 \int_0^\infty dq \frac{(q+p_1) \left [
\psi(p_1, p_2) - \psi(q, p_2)e^{i(p_1-q)x^-/2}\right ]}{2\sqrt{p_1q}(q-p_1)^2}\\
+ G^2 \int_0^\infty dq \frac{(q+p_2)\left [
\psi(p_1, p_2) - \psi(p_1, q)e^{i(q-p_2)x^-/2}\right ]}{2\sqrt{p_2q}(q-p_1)^2}\\
+ G^2 \int_0^{p_1+p_2} dq \frac{(q+p_1)(p_1+2p_2-q)} 
{4\sqrt{p_1p_2q(p_1+p_2-q)}(q-p_1)^2}\\
\times \left [ \psi(p_1, p_2) - \psi(q, p_2)\right ]\\
+ G^2 \frac{\pi}{4\sqrt{p_1p_2}} \psi(p_1,p_2)\\
+\lambda_1\int_0^{p_1+p_2} dq\frac{\psi(p_1, p_2) +
\psi(q,p_2)}{\sqrt{p_1p_2q(p_1+p_2-q)}}\\
+\lambda_c \int_0^{p_1+p_2}dq \frac{ \psi(p_1,p_2) }{v^+(p_1p_2)}\\
+\lambda_c\int_0^{p_1+p_2} dq \frac{\psi(q,p_2)e^{i(q-p_2)x^-/2}}{v^+
\sqrt{p_1p_2q(p_1+p_2-q)}}\\
+\lambda_c\int_0^{p_1+p_2} dq \frac{\psi(p_1,q)e^{i(p_1-q)x^-/2}}{v^+
\sqrt{p_1p_2q(p_1+p_2-q)}}
\end{eqnarray*}

The momentum
cut-off, along with the ``plus'' component of the four-velocity of
the quark/anti-quark pair ($v^+$), must be carefully extrapolated to
infinity. By considering the asymptotic dependence of the momentum-space
wave function in our eigenvalue equation, this dependence of the energy on the
cut-off, $\Lambda$, is found to be,\\

\begin{eqnarray}
E \sim -\frac{1}{\Lambda^4}
\end{eqnarray}

The dependence on $v^+$ is assumed to be of the form,\\

\begin{eqnarray}
E \sim \left ( \frac{1}{v^{+\ 2}} \right )^p
\end{eqnarray}

$p$ is found empirically from the data for each configuration. Typically,
$p$ is around $0.20$ when extrapolating the lowest energy eigenvalue 
corresponding to the ground state $Q\bar{Q}$ Potential.
 
As was discussed in previous work \cite{us}, the $Q\bar{Q}$
Potential is an important observable for us to study. One of the reasons is
the lack of manifest rotational invariance in
Light Front field theories\cite{ji}.
Because rotational invariance is not a manifest symmetry of the Light
Front hamiltonian, it is sensitive to the the dynamics included in the
hamiltonian. If the dynamics of the hamiltonian are incorrect, rotational
invariance will not be observed and this will be obvious from the shape of
the $Q\bar{Q}$ Potential. From Euclidean Lattice Monte
Carlo Simulations \cite{morningstar}
the $Q\bar{Q}$ Potential is
known to be a linear function of separation with a slope, $\sigma$, known
as the {\it string tension}. If the $Q\bar{Q}$ Potential were to depend on
orientation of the quarks (i.e. {\bf not} rotationally invariant) then
the string tension would be different for these different orientations. In
such a case the $Q\bar{Q}$ Potential would not be a very good line:
the points would be scattered in a wedge between the maximum and minimum
string tensions.\\

In a full 3+1 dimensional calculation,
the five couplings of the effective action would be tuned independently
to obtain a rotationally invariant $Q\bar{Q}$ Potential. The five 
couplings are $\mu^2$ (the link field mass), $\lambda_1$, $\lambda_2$, 
$\lambda_p$, and $\lambda_c$.
In this 2+1 dimensional calculation, the plaquette interaction
and the interaction between links that turn a corner
are superfluous since there is only one discrete direction. Thus there are
only three free parameters to tune in the hamiltonian: $\mu^2$, $\lambda_1$,
and $\lambda_c$. We found empirically that adjusting $\mu^2$ and $\lambda_1$
affected the global rotational invariance of the points while
adjusting the contact coupling, $\lambda_c$, most affected the intercept
of the $Q\bar{Q}$ Potential. As was shown in our previous work on the
2+1 Dimensional $Q\bar{Q}$ Potential, its slope in our units
should be $G^2\frac{\pi}{2}$.\\

The strong coupling limit provides us with an important point of comparison
for our data. The strong coupling limit does {\bf not} exhibit rotational
invariance. In this limit ($\mu^2\rightarrow \infty$), the $Q\bar{Q}$ Potential
has the form,

\begin{eqnarray}
V(x_\perp, x_L) = G^2\frac{\pi}{2} \left ( |x_\perp| + |x_L| \right )
\end{eqnarray}

The data
points in this limit would be scattered within a wedge. The lower slope
would be $G^2\frac{\pi}{2}$ and the upper slope would be
$\sqrt{2}G^2\frac{\pi}{2}$, corresponding to a separation which is
``diagonal'', having equal transverse and longitudinal displacements. If
the data points in our calculation stay close to the lower slope and well 
away from the upper slope indicated by the strong coupling limit, rotational
invariance can be claimed.
\section{Results and Analysis}
\begin{figure}
\includegraphics[scale=0.50]{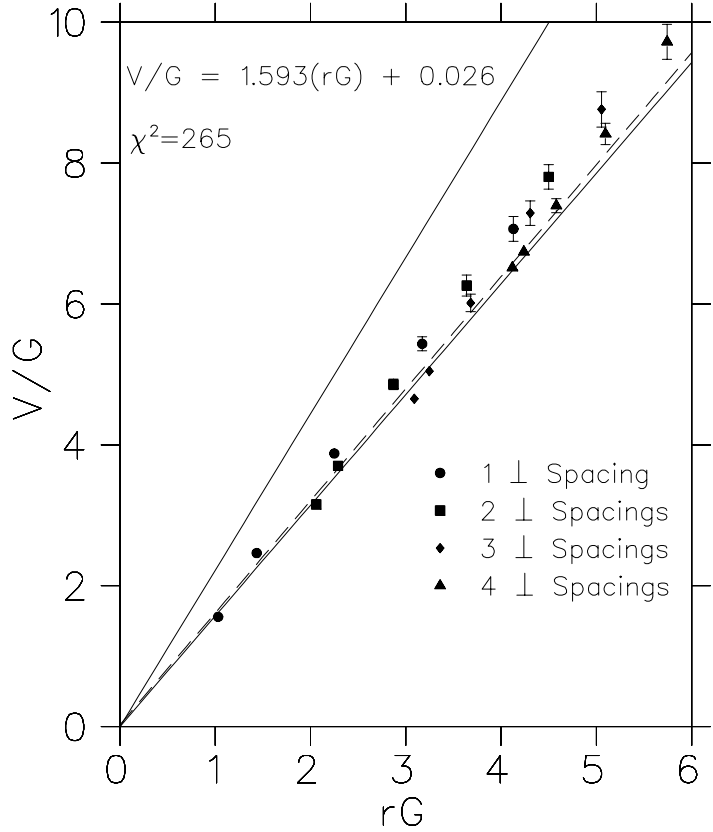}
\caption{Plot of the poorly tuned $Q\bar{Q}$ Potential. The solid lines
represent the maximum and minimum slopes for the strong coupling limit. The
dashed line is the best fit through the data points. $\chi^2$ for the fit
is 265. $\chi^2$ per degree of freedom is 13.3\label{baddata}}
\end{figure}

\begin{figure}
\includegraphics[scale=0.50]{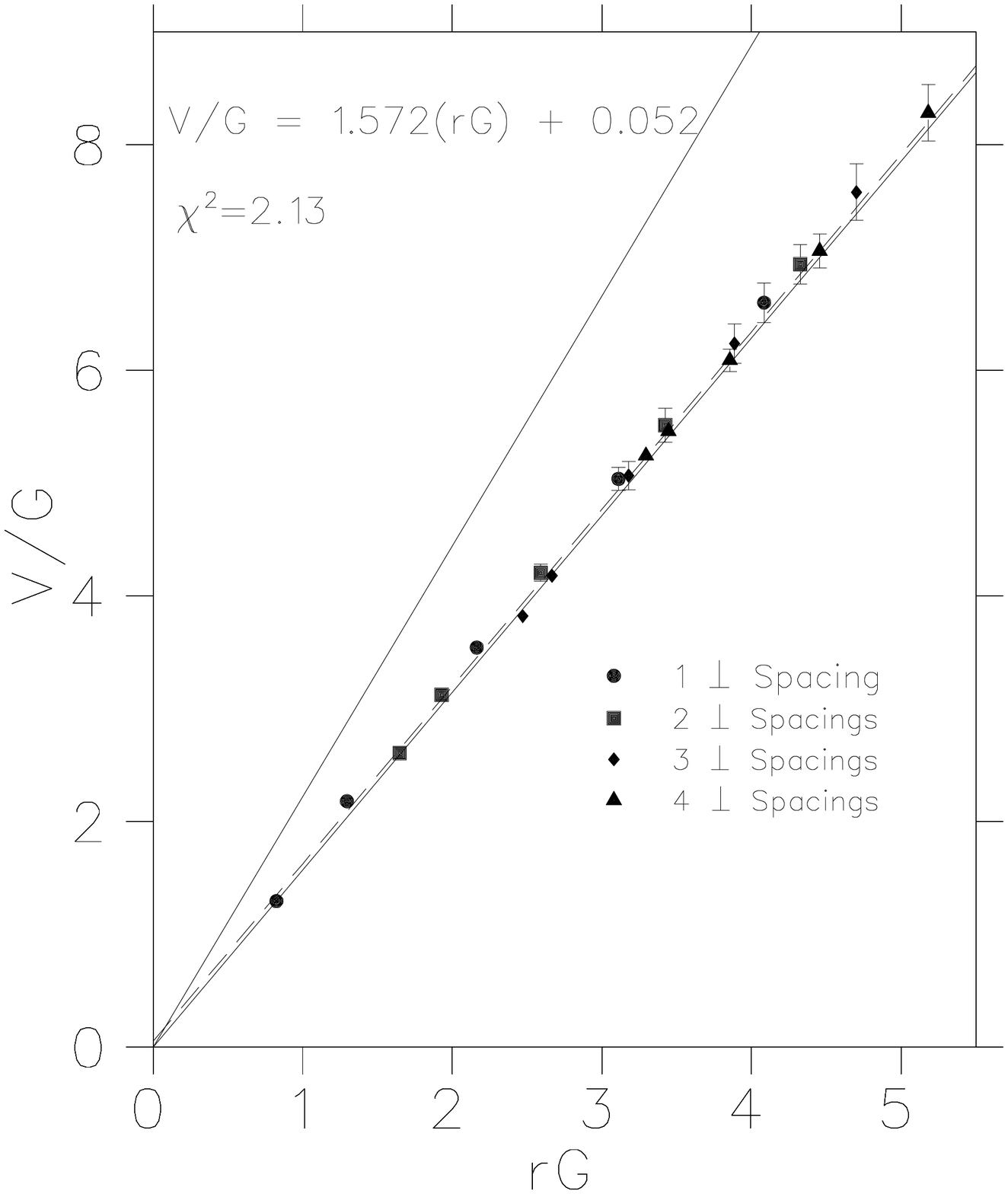}
\caption{Plot of the finely tuned $Q\bar{Q}$ Potential. The solid lines
represent the maximum and minimum slopes for the strong coupling limit. The
dashed line is the best fit through the data points. $\chi^2$ for the fit
is 2.13. $\chi^2$ per degree of freedom is 0.11.\label{gooddata}}
\end{figure}

Plots of the $Q\bar{Q}$ Potential data are shown in Figures \ref{baddata} and
\ref{gooddata}. In Figure \ref{baddata}
a set of data is shown which has some anisotropy.
While it does not exhibit the uniform distribution of points characteristic
of the strong coupling limit, the data points do not all lie upon the
ideal line.
Figure \ref{gooddata}
represents the best-tuned data for the $Q\bar{Q}$ Potential that we obtained.
$\chi^2$-analysis shows that this line has a slope very close
to the expected slope of $G^2\frac{\pi}{2}$. The rotational invariance
of this data is quite striking, thus we are very confident in the
dynamics that were included in this incarnation of the Color Dielectric
formulation of the Transverse Lattice. 

The Tranverse Lattice is essentially an effective theory, but one which
is grounded in the fundamental underlying dynamics of the point-like
gluon fields. While no appeal has been made in this work to actually
calculate the relationship between the link fields on the localized gluonic
degrees of freedom, such a study is possible in principle. Determining this
relationship would lend a rigor to this effective theory  
that most phenomenological models do not enjoy. Such a calculation could be
performed in a pure glue (i.e. quenched)
Euclidean Lattice Monte Carlo Simulation where a smearing transformation is
performed on the Euclidean Lattice link
fields. Determining the couplings for the effective action in this
formalism holds great appeal since it does not rely on tuning them to 
fit data. Instead, the calculations could be used to test the validity of
the theory or as predictions in their own right.\\
 
The couplings found from this study form a foundation on which to begin
the more realistic 3+1 dimensional calculations. In 3+1 dimensions,
the plaquette coupling would need to be included along with a new coupling
similar to $\lambda_1$, but acting on two links which turn a corner rather than
go in a straight line. Work on the 3+1 dimensional $Q\bar{Q}$ Potential is
currently underway by the authors. As computer storage becomes more affordable,
another extension to this work can be considered: relaxing the minimal string
approximation. Including such states would allow us to test the validity of
the approximation as well as to relax the restriction on the Fock space.
Including such transverse oscillations of the link field string could have
important contributions for the excited state potentials.
In addition, the excited state potentials themselves can also be
extracted from the data. These potentials correspond to an excited gluonic
string and can be interpreted as the potential experienced by the valence
quarks in hybrid mesons. Comparing the excited state potentials to
phenomenological models such as the Flux Tube Model \cite{isgur}
will give insight into
the phenomenology represented by these models. The suggested extensions
to the present work, as stated above, demonstrate only a small fraction of
the utility of the Transverse Lattice. 

\bibliography{2dpot0902}

\end{document}